\documentclass[aps,english,superscriptaddress,preprintnumbers,reprint,footinbib,amsmath,amssymb,prl]{revtex4-2}
\usepackage[version=3]{mhchem}
\usepackage{graphicx}
\usepackage{dcolumn}
\usepackage{bm}
\usepackage{hyperref}
\hypersetup{
    colorlinks=true,
    linkcolor=blue,
    filecolor=blue,
    urlcolor=blue,
   citecolor=blue,
}
\usepackage{enumitem}

\begin{document}


\title{Two-dimensional chiral waveguide quantum electrodynamics: long range qubit correlations and flat-band dark polaritons}

\author{Y. Marques}
\email[corresponding author:]{ yurimarques111@gmail.com}
\affiliation{ITMO University, St.~Petersburg 197101, Russia}

\author{I. A. Shelykh}
\affiliation{ITMO University, St.~Petersburg 197101, Russia}
\affiliation{Science Institute, University of Iceland, Dunhagi-3, IS-107,
Reykjavik, Iceland}

\author{I. V. Iorsh}
\email[corresponding author:]{ iorsh86@yandex.ru}
\affiliation{ITMO University, St.~Petersburg 197101, Russia}

\date{\today}

\begin{abstract} 
We consider a two-dimensional extension of the 1D waveguide quantum electrodynamics and investigate the nature of linear excitations in two-dimensional arrays of qubits coupled to networks of chiral waveguides. We show that the combined effects of chirality and long-range photon mediated qubit-qubit interactions lead to the emergence of the two-dimensional flat bands in the polaritonic spectrum, corresponding to slow strongly correlated light.
\end{abstract}

\maketitle

\section{Introduction}

Arrays of quantum emitters, placed in the vicinity of photonic nanostructures are now in focus of an intensive research, and constitute the field of Waveguide Quantum Electrodynamics (WQED)~\cite{Roy2017,KimbleRMP2018}. While specific material platforms of WQED span from artificial arrays of cold atoms~\cite{Corzo2019} to radio-frequency circuits with superconducting qubits~\cite{vanLoo2013,Mirhosseini2019}, all of them possess a common feature, namely the long range inter-qubit interaction, mediated by the exchange of the propagating photons. The onset of long-range correlations enables to access the plethora of the intriguing physical phenomena such as emergence of unconventional topological phases~\cite{kim2020quantum,Chang2020}, 
collective super-radiance and sub-radiance ~\cite{Ke2019,kornovan2019extremely,Zhang2019arXiv,Albrecht2019,Henriet2019,Zhang2019,Ke2019}, and paves way towards design and engineering of prospective quantum networks~\cite{kimble2008quantum}.

One of the most prominent consequences of the long range inter-qubit interaction is peculiar physics of the multiparticle bound states, emerging in the WQED structures~\cite{Poddubny2020,Zhang2020,PhysRevX.10.031011}. They are related to unconventional dispersion of the polaritons, hybrid quasi-particles formed in the system due to the coherent mixture of propagating waveguide photons and qubit excitations. Specifically, the presence of the long-range correlations in the system makes possible the appearance of quasi-flat polariton dispersion, which results in emergence of the unconventional few-photon bound states, absent in lattice models with nearest neighbour interactions. The study of the dispersion properties of the polaritons in WQED is thus an important task, crucial for understanding of the nature of unconventional multiphotonic states in the considered systems.

Chirality is an immanent property of many of WQED geometries~\cite{Lodahl2017}. In arrays of cold atoms or semiconductor quantum dots coupled to photonic waveguides, chirality arises due to the spin-momentum locking, immanent for  the confined electromagnetic modes in these structures~\cite{bliokh2015quantum,coles2016chirality} and results in the decay of a certain transition in the emitter to the polarized photonic mode propagating in a certain direction~\cite{Pichler2015}. Alternatively, chirality may be realized in systems, based on topological waveguides, formed at the edges of a photonic topological insulator~\cite{PhysRevB.101.205303, mehrabad2019chiral}.

Up to now, most of the WQED research was focused on  one-dimensional geometries. At the same time, two-dimensional systems, based e.g. on arrays of atoms coupled to the modes of photonic crystals or metasurfaces are certainly of interest, since they can be used as bosonic simulators of the correlated phases in condensed matter ~\cite{Gonzalez-Tudela2015} and have been recently realized experimentally~\cite{Goban2014}.

\begin{figure}[t]
	\centerline{\includegraphics[width=3.0in,keepaspectratio]{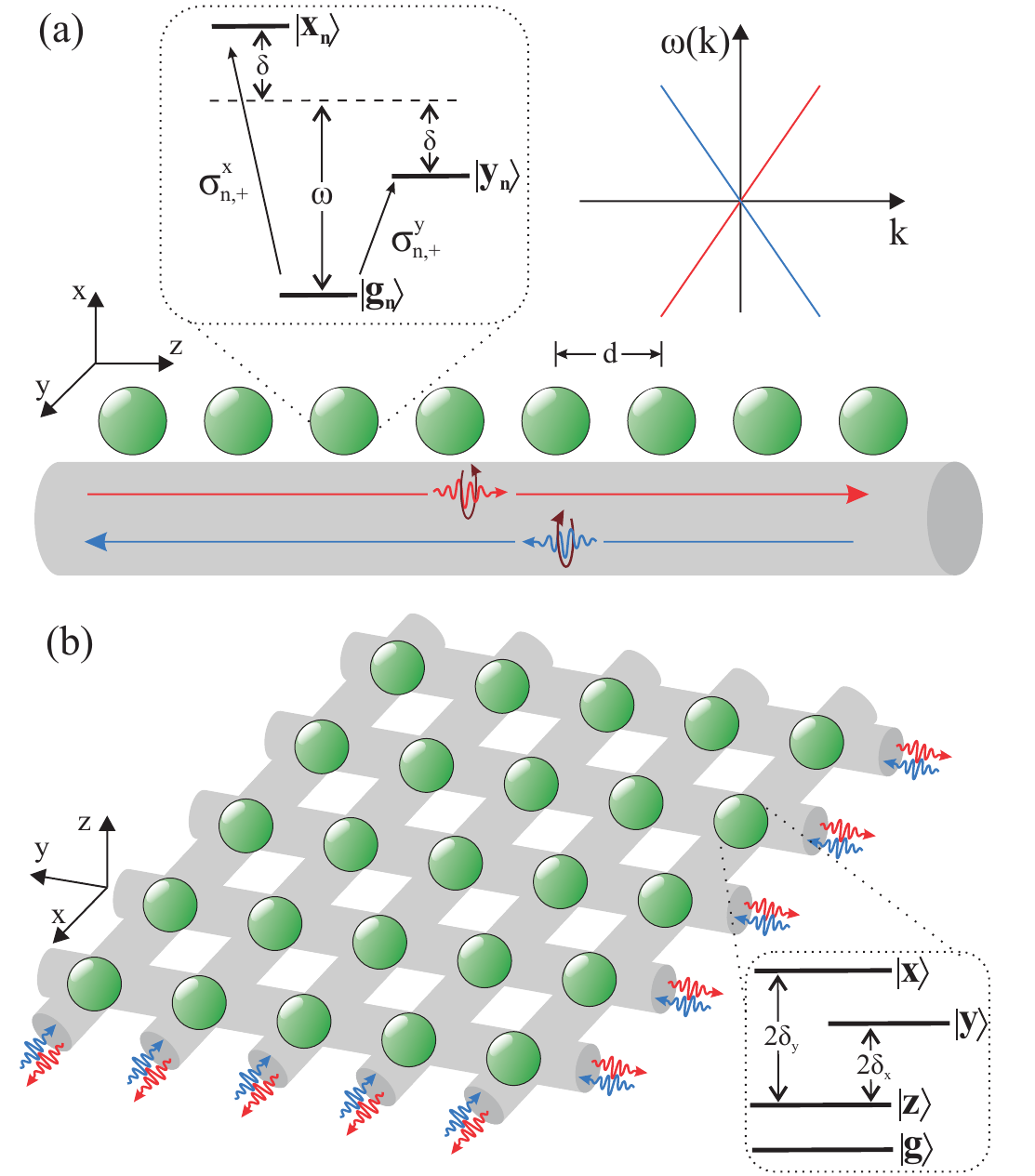}}
	\caption{\label{fig:Fig1}(a) Sketch of the considered setup consisting of an $N$-atomic chain placed over a waveguide supporting counter propagating modes with opposite circular polarizations. Three-level atoms in V-type configuration with dipole excited states $\bigl|x\bigr\rangle$ and $\bigl|y\bigr\rangle$, detuned by $2\delta$ from each other are chirally coupled to right- and left-propagating modes $\bigl|R\bigr\rangle$ and $\bigl|L\bigr\rangle$  with linear dispersion relation $\omega(k)$. (b) Two-dimensional square lattice of atoms placed over a waveguide lattice hosting a pair of circularly polarized modes propagating along x and another pair propagating along y. In the network of chiral waveguides, the atoms acquire an extra linearly polarized dipole transition along z with $\bigl|y\bigr\rangle$ and $\bigl|z\bigr\rangle$ detuned by $2\delta_{x}$ and $\bigl|x\bigr\rangle$ and $\bigl|z\bigr\rangle$ detuned by $2\delta_{y}$.} \label{fig:1}
\end{figure}

In this work, we consider the systems, consisting on one or two-dimensional arrays of atoms coupled to a single chiral waveguide or a network of chiral waveguides, as it is shown  schematically in Fig.~\ref{fig:1}. Each waveguide is characterized by the spin-momentum locking: two orthogonal circularly polarized modes propagate in the opposite directions. In the basis of linear polarization, each atom in the one-dimensional array has two polarized dipole transitions aligned along $x$ and $y$, while in the two-dimensional lattice atoms present an extra linearly polarized transition aligned along $z$. When these transitions are degenerate, the 1D system is purely chiral with left and right circularly polarized photons propagating in specific directions. Once a detuning $2\delta$ is introduced between the transition frequencies, which can be realized by application of a properly aligned stationary electric field resulting in the Stark effect, the circularly polarized states become mixed. In the 2D lattice, the left and right circularly polarized photons naturally become mixed even when the transitions are degenerate, and thus, leads to a radical reshaping of the dispersions of the polariton modes and in certain conditions it results in the emergence of the polariton flat bands, corresponding to slow strongly correlated light.

\section{one-dimensional chain model}

We start from the consideration of the 1D case. The corresponding geometry is schematically illustrated in Fig.~\ref{fig:Fig1}. It consists of N regularly-spaced identical three-level atoms placed over a 1D chiral waveguide. The corresponding Hamiltonian, within the rotating wave approximation, reads:
\begin{eqnarray}
\mathcal{H} & = & \sum_{k,\lambda}\hbar\omega_{k,\lambda}a_{k,\lambda}^{\dagger}a_{k,\lambda} + \sum_{n,\nu}\hbar\omega_{\nu}\sigma_{n,+}^{\nu}\sigma_{n,-}^{\nu}\nonumber \\
 & + & \frac{\hbar g}{\sqrt{2l}}\sum_{k,n}\left(e^{ikz_{n}}\sigma_{n,+}^{x}-ie^{ikz_{n}}\sigma_{n,+}^{y}\right)a_{k,R} + \text{h.c.}\nonumber \\
 & + & \frac{\hbar g}{\sqrt{2l}}\sum_{k,n}\left(e^{ikz_{n}}\sigma_{n,+}^{x}+ie^{ikz_{n}}\sigma_{n,+}^{y}\right)a_{k,L} + \text{h.c.},
 \label{eq:H}
\end{eqnarray}
wherein $a_{k,\lambda}^{\dagger}(a_{k,\lambda})$ represent the creation (annihilation) operator of waveguide photons, characterized by propagation direction $\lambda=R,L$ which is locked with right and left circular polarization, wavenumber $k$ and linear dispersion $\omega_{k,\lambda}(k)=\pm vk$ (see Fig.~\ref{fig:Fig1}(a)). The three-level atoms positioned at $z_{n}$, are characterized by two liner polarized detuned dipole transitions, $\nu=x,y$ corresponding to frequencies $\omega_{\nu}=\omega \pm \delta$ and raising and lowering operators $\sigma_{n,+}^{\nu} = \bigl|\nu_{n}\bigr\rangle\bigl\langle g_{n}\bigr|$, $\sigma_{n,-}^{\nu} = \bigl|g_{n}\bigr\rangle\bigl\langle \nu_{n}\bigr|$. The parameters $g$ and $l$ stand for the interaction constant and the normalization length, respectively.
In order to effectively decouple the atomic subsystem from the photonic modes, which corresponds to the use of Markovian approximation valid for $\delta,g \ll \omega$, we apply Schrieffer-Wolff transformation~\cite{SW-Kondo,SW-DL}, performing the unitary transformation $\mathcal{H}_{\textrm{eff}}  = e^{-S}\mathcal{H}e^{S}$, with
\begin{eqnarray}
S & = & \frac{g}{\sqrt{2l}}\sum_{k,n}\bigl(\frac{\sigma_{n,-}^{x}a_{k,L}^{\dagger}}{\omega_{k,L}-\omega_{x}}+\frac{\sigma_{n,-}^{x}a_{k,R}^{\dagger}}{\omega_{k,R}-\omega_{x}}\bigr)e^{-ikz_{n}}-\text{h.c.}\nonumber \\
 & + & \frac{ig}{\sqrt{2l}}\sum_{k,n}\bigl(\frac{\sigma_{n,-}^{y}a_{k,R}^{\dagger}}{\omega_{k,R}-\omega_{y}}-\frac{\sigma_{n,-}^{y}a_{k,L}^{\dagger}}{\omega_{k,L}-\omega_{y}}\bigr)e^{-ikz_{n}}-\text{h.c.} \label{eq:S}
\end{eqnarray}
being the anti-Hermitian generator, and writing the effective Hamiltonian in the form $\mathcal{H}_{\textrm{eff}}=H_{0}+\frac{1}{2}\left[H_{v},S\right]$, wherein $H_{0}$ and $H_{v}$ represent the diagonal and off-diagonal parts of (\ref{eq:H}). The detailed Schrieffer-Wolff approach is presented in Supplemental Material (S.1).

\begin{figure}[t]
	\centerline{\includegraphics[width=3.4in,keepaspectratio]{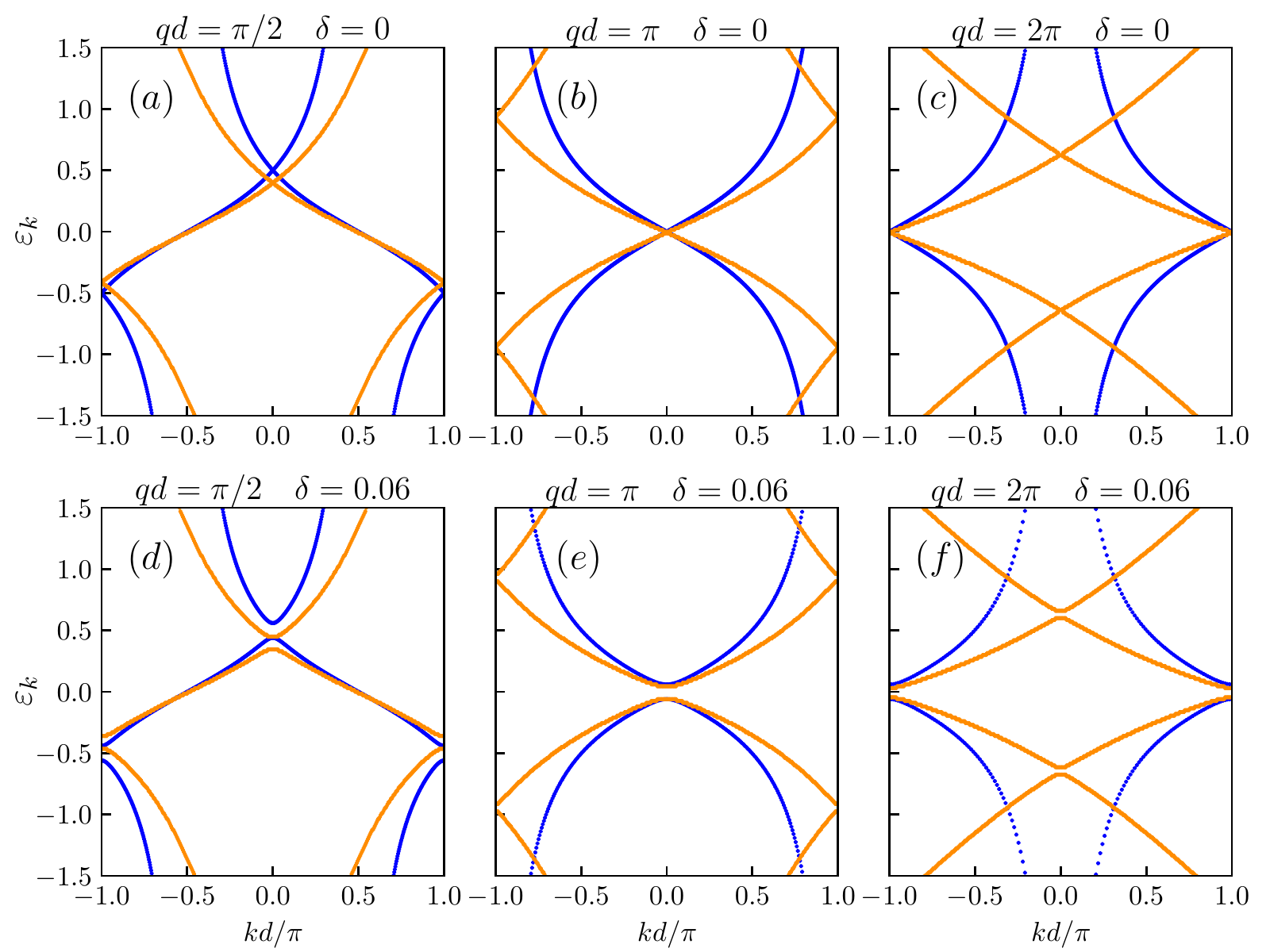}}
	\caption{Polariton dispersion for one-dimensional chain obtained by using Schrieffer-Wolff transformation (blue) in comparison with the exact solution (orange). The upper panels correspond to the case of zero detuning with dispersion branches corresponding to the right (R) and left (L) propagating modes. The lower panels illustrate the case of a finite detuning $\delta$, when R and L modes become mixed. \label{fig:Fig2}}
\end{figure}

\begin{figure}[t]
    \centerline{\includegraphics[width=3.4in,keepaspectratio]{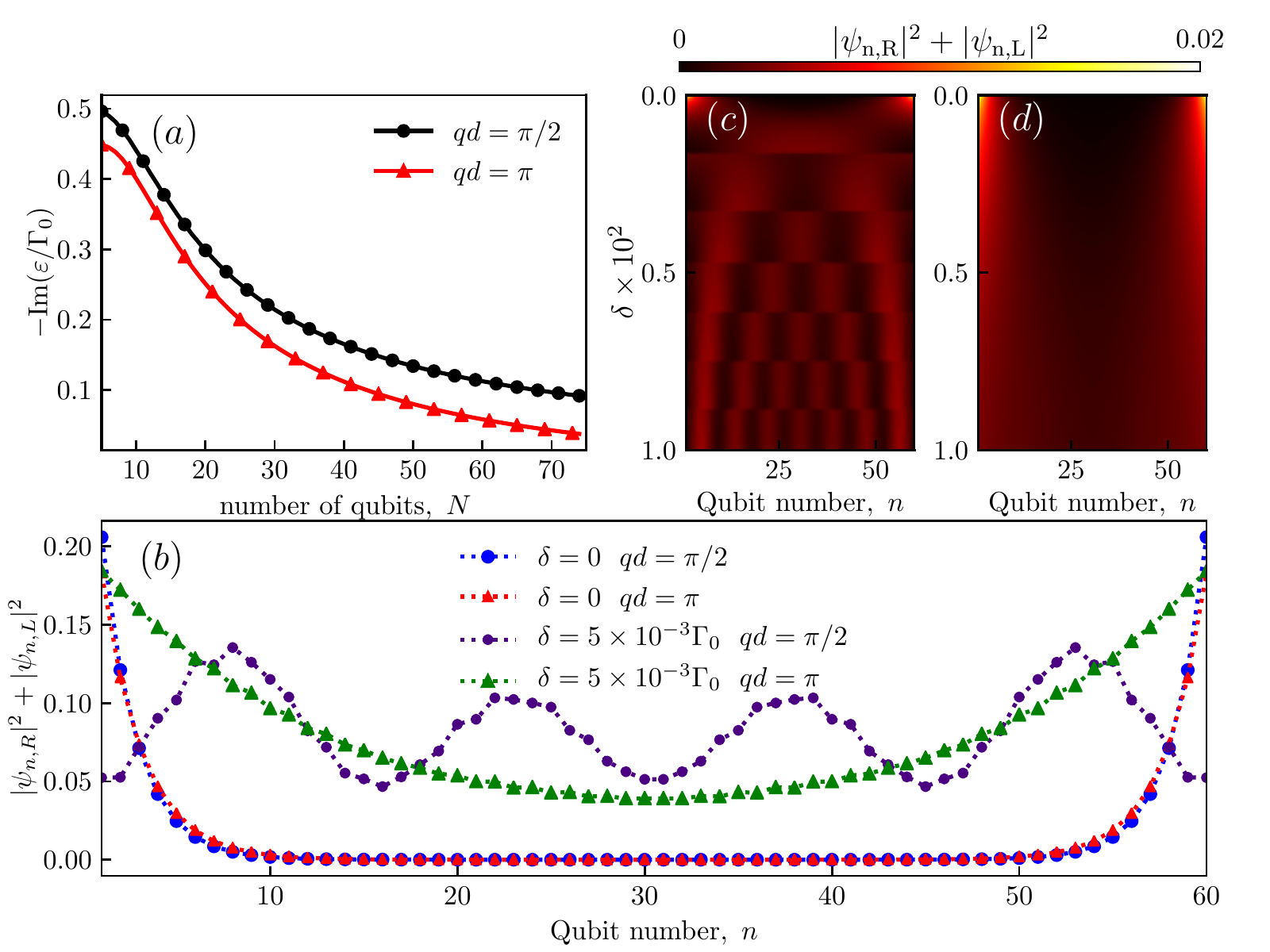}}
    \caption{(a) Radiative decay rate profile scaled to $\Gamma_0$, $-\textrm{Im}(\varepsilon/\Gamma_{0})$, as a function of the array size N for $qd=\pi/2$ and $qd=\pi$. For a better viewing, degenerate curves are slightly shifted. (b) Photonic distribution corresponding to the darkest state for an array composed by $N=60$ atoms. Panels (c) and (d) display illustrate the effect of detuning $\delta$ on photonic distribution  for $qd=\pi/2$ and $qd=\pi$, respectively.
    \label{fig:Fig3}}
\end{figure}

As a result, the effective real-space Hamiltonian of the system, written in terms of the qubit operators only, reads:
\begin{eqnarray}
\mathcal{H}_{\textrm{eff}} & = & \sum_{n=1}^{N}\delta\bigl(b_{n,R}^{\dagger}b_{n,L}+\textrm{h.c.}\bigr) -i\frac{\Gamma_{0}}{2}\sum_{m,n=1}^{N}\Theta(m-n) \nonumber \\
& \times & e^{iqd|m-n|} \bigl(b_{m,R}^{\dagger}b_{n,R}+b_{n,L}^{\dagger}b_{m,L}\bigr),
\label{eq:Heff}
\end{eqnarray}
where the individual decay rate of a single atom $\Gamma_{0}= g^{2}/v$ defines characteristic energy scale of the system, $q=\omega/v$ is the wavevector of the photon, mediating the interaction between qubits equidistantly spaced by $d$, and $\Theta$ is the Heaviside step function defined within
the half-maximum convention. We also used bosonisation procedure for qubits,  replacing $\sigma_{n,+}^{x,y}$ by regular bosonic operator in the basis of circular polarization $b_{n,x}^{\dagger}=(b_{n,R}^{\dagger}+b_{n,L}^{\dagger})/\sqrt{2}$ and $ b_{n,y}^{\dagger}=i(b_{n,R}^{\dagger}-b_{n,L}^{\dagger})/\sqrt{2}$, as we are interested in single-excitation states only. Note that the Hamiltonian accounts for excitations with unidirectional interaction, in which an excitation of qubit $m$ with right polarization is transferred only to qubits on its right ($b_{m,R}^{\dagger}b_{n,R}$) and excitations with left polarization are transferred in the opposite direction ($b_{n,L}^{\dagger}b_{m,L}$). The interaction between these counter-propagating excitations emerges as the on-site interaction mediated by the detuning $\delta$. It is worth mentioning that by integrating out the infinite degrees of freedom of the waveguide reservoir, the non-Hermitian Hamiltonian (\ref{eq:Heff}) effectively describes an open system of interacting qubits, in which its real and imaginary part accounts for coherent and dissipative interaction between qubits, respectively.
In order to explore the outcome of chiral propagation in the system dispersion relation, it is instructive to move from the finite lattice to the infinite system in k-space via Fourier transform as detailed in supplemental material (S.2). As a result, the bilinear Hamiltonian in k-space reads
\begin{eqnarray}
\mathcal{H}_{\textrm{eff}} & = &  \frac{\Gamma_{0}}{4} \sum_{k} \textrm{cot}\left(\frac{qd - kd}{2}\right)  b_{k,R}^{\dagger}b_{k,R}\nonumber\\
& + & \frac{\Gamma_{0}}{4}\sum_{k}\textrm{cot}\left(\frac{qd + kd}{2}\right) b_{k,L}^{\dagger}b_{k,L} \nonumber\\
& + & \delta(b_{k,R}^{\dagger}b_{k,L}+\mathrm{h.c.}).\label{eq:Heff-k}
\end{eqnarray}
The translational invariance of the infinite system provides an Hermitian Hamiltonian obeying the Bloch's theorem, in which its direct diagonalization gives dispersion curves, shown in  Fig.~\ref{fig:Fig2}. Their shape is determined by the phase $qd$ and detuning $\delta$. For the case of $\delta=0$ illustrated by the upper panels, $R$ and $L$ modes are completely decoupled. The anti-Bragg structure dispersion with $qd=\pi/2$ for finite $\delta$ presents a middle slow polariton branch gapped from the upper and lower fast polariton bands (see Fig.~\ref{fig:Fig2} (d)). For Bragg structures, $qd=\pi$ and $qd=2\pi$, $\delta$ is responsible for open a gap around $\varepsilon_{k}=0$ as can be seen in Figs.~\ref{fig:Fig2} (e,f).
As a consequence of Markov approximation, the polariton dispersion present nonphysical regions with infinite group velocities for any phase $qd$. Specifically for $qd=2\pi$ and $qd=\pi$ these divergences coincides with regions of interest corresponding respectively to the center and edges of the first Brillouin zone, and thus, can be clarified by the exact solution obtained via the transfer matrix method.
The transfer matrix is a powerful approach to obtain the lattice dispersion relation for single-photon propagation in one-dimensional infinite lattices, once that it precisely accounts the field scattering throughout the array~\cite{Deutsch1995,joannopoulos2011book,Asenjo2017}. Hence, for the considered setup the transfer matrix over the single period of the structure composed by a layer of thickness
$d$ intersected by a scatterer positioned at $z=0$ reads:
\begin{eqnarray}
\mathcal{T}=\frac{1}{t}\left(\begin{array}{cc}
\left(t^{2}-r^{2}\right)e^{iqd} & r\\
-r & e^{-iqd}
\end{array}\right) & ; & \ \ t=1+r,\label{eq:T0}
\end{eqnarray}
wherein $r$ and $t$ stand for amplitude reflection and transmission coefficients of the scatterer, respectively. The matrix specifying the pair of discrete levels at $\omega_x$ and $\omega_y$ of the scatterer, written in the circularly polarized basis (see Eq.~(S22)), reads
\begin{eqnarray}
\hat{h} & = &
\frac{2i\Gamma_{0}}{(\omega-\varepsilon_{k})^{2}-\delta^{2}}
\left(\begin{array}{cc}
\omega-\varepsilon_{k} & -\delta\\
-\delta & \omega-\varepsilon_{k}
\end{array}\right).
\end{eqnarray}
Following the Green's function formalism~\cite{economou2013}, the amplitude transmission coefficient can be found via\begin{eqnarray}
t & = & 1+\left\{ \hat{h}\times\left[\hat{I}-\hat{G}_{0}(0)\hat{h}\right]^{-1}\right\} _{11}
\end{eqnarray}
with the Green's function of the scatterer written as
\begin{eqnarray}
\hat{G}_{0}(z) & = & \left(\begin{array}{cc}
\Theta(z)e^{ik|z|} & 0\\
0 & \Theta(-z)e^{-ik|z|}
\end{array}\right).
\end{eqnarray}
Therefore, the dispersion equation for the eigenmodes $\cos(kd)=\frac{1}{2}\mathrm{Tr}(\mathcal{T})$, detailed in the supplemental material (S.3), reads
\begin{eqnarray}
\cos(kd) & = & \cos(qd)\left[\frac{\left(\omega-\varepsilon_{k}\right)^{2}-\delta^{2}-\Gamma_{0}^{2}/4}{\left(\omega-\varepsilon_{k}\right)^{2}-\delta^{2}+\Gamma_{0}^{2}/4}\right] \nonumber \\
& - & \sin(qd)\left[\frac{\left(\omega-\varepsilon_{k}\right)\Gamma_{0}}{\left(\omega-\varepsilon_{k}\right)^{2}-\delta^{2}+\Gamma_{0}^{2}/4}\right].
\end{eqnarray}
with its eigenvalues $\varepsilon_{k}$ shown in Fig.~\ref{fig:Fig2} for different phases $qd$. One can see that the exact solution can diverge significantly from those obtained within Markov approximation. In particular, it does not contain nonphysical regions with infinite group velocity around $k=0$ for $qd=2\pi$, where additional gap is opened instead, and substantially modifies band structure in the vicinity of the band edges $k=\pm\pi/d$ for the case $qd=\pi$.
For finite structures, the effective Hamiltonian (\ref{eq:Heff}) is characterized by complex eigenenergies $\varepsilon$, with real part corresponding to a frequency shift relative to $\omega$, and imaginary part corresponding to radiative decay rate,  $\Gamma_{1D}=-\textrm{Im}(\varepsilon)$ \cite{Yongguan2019}. The states can be classified into superradiant ($\Gamma_{1D} \sim N\Gamma_{0}$), bright ($\Gamma_{1D} \sim \Gamma_{0}$), and
subradiant ($\Gamma_{1D} \ll \Gamma_{0}$). Noteworthy, the dependence of the darkest state decay rate on the number of atoms for the considered setup, presented in Fig.~\ref{fig:Fig3} (a), reveals a scaling $\Gamma_{1D}\sim N^{-1}$ in contrast to  $\Gamma_{1D}\sim N^{-3}$ scaling, characteristic for 1D waveguides hosting linear polarized modes. This is in full accordance with the results of~\cite{PhysRevB.96.115162}, where $N^{-1}$ of the subradiant modes for chiral waveguides has been predicted.
The subradiant states correspond to the lowest group velocity modes in the dispersion relation, so that subradiant states for $qd=\pi/2$ are related to the modes found at the edges and center of the first Brillouin zone, while for $qd=\pi$ they correspond just to the modes in the vicinity of $k=0$. The phase $qd=2\pi$ is disregarded due to the nonphysical regime around $k=0$ and its substantial divergence in relation to the exact solution.
 
 Aiming to analyze the role of photonic distribution in chiral propagation along the array in single-excitation regime, we define the occupation basis $\bigl|\Psi\bigr\rangle=\sum_{n,\lambda}\psi_{n,\lambda}b_{n,\lambda}^{\dagger}\bigl|0\bigr\rangle$, with $\psi_{n,\lambda}$ being the probability amplitude to find an excitation at the site $n$ with polarization $\lambda$. The photonic distribution corresponding to the darkest state displayed in Fig.~\ref{fig:Fig3} (b) reveals strong localization at the edges of the array in which the right (left) polarized mode is found at the right (left) end of the array for zero detuning and delocalized profiles for finite detunings. The smooth crossover from strongly localized to delocalized distributions is well perceived in Figs.~\ref{fig:Fig3} (c,d).

\section{two-dimensional lattice model}

In the two dimensional (2D) model, we consider a $N \times N$ square lattice of atoms at the nodes of a quadratic waveguide composed by a set of horizontal and vertical chiral waveguides in $xy$ plane as depicted in Fig.~\ref{fig:1} (b).
\begin{figure}[t]
\centerline{\includegraphics[width=3.4in,keepaspectratio]{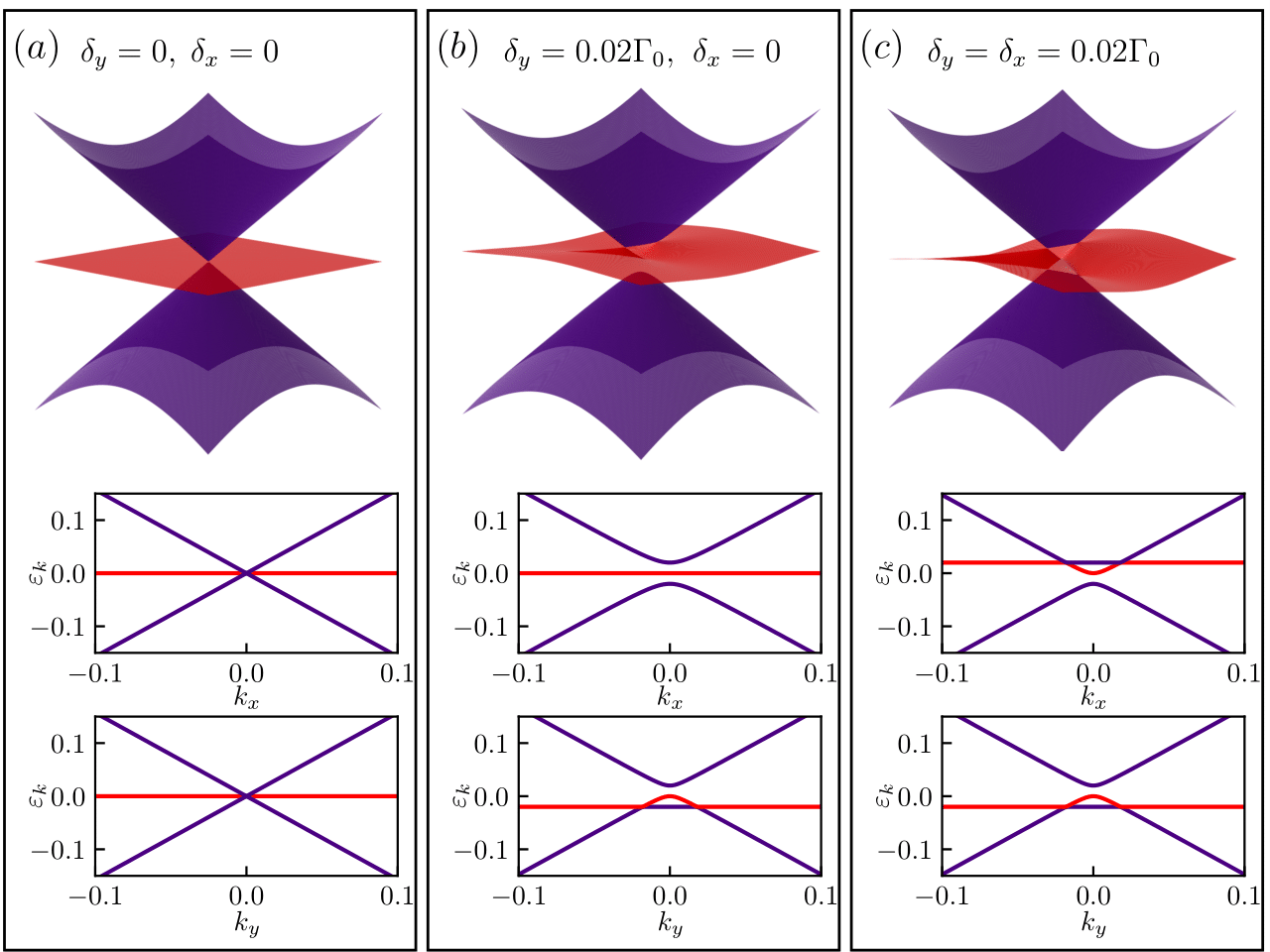}}
    \caption{Dispersions in 2D system corresponding to low-energy effective Hamiltonian  (\ref{eq:H2d-k}) and their projections corresponding to $k_{x}=0$ (middle panels) and $k_{y}=0$ (lower panels) for (a) $\delta_{x}=\delta_{y}=0$, (b) $\delta_{x}=0$ and $\delta_{y}=0.02\Gamma_{0}$, and (c) $\delta_{x}=\delta_{y}=0.02\Gamma_{0}$.
    \label{fig:Fig4}}
\end{figure}
Separately, waveguides aligned along x direction excites dipole transitions $\bigl|R_{yz}\bigr\rangle = (\bigl|y\bigr\rangle + i\bigl|z\bigr\rangle)/\sqrt{2}$ and $\bigl|L_{yz}\bigr\rangle = (\bigl|y\bigr\rangle - i\bigl|z\bigr\rangle)/\sqrt{2}$ in each qubit, while the waveguides along y direction excites transitions $\bigl|R_{zx}\bigr\rangle = (\bigl|z\bigr\rangle + i\bigl|x\bigr\rangle)/\sqrt{2}$ and $\bigl|L_{zx}\bigr\rangle = (\bigl|z\bigr\rangle - i\bigl|x\bigr\rangle)/\sqrt{2}$. Therefore, when they are together forming the 2D network, each qubit has linear dipole transitions along x,y and z, in which the detuning frequency between the corresponding energy levels are $2\delta_{x}= \omega_{y}-\omega_{z}$ and $2\delta_{y}= \omega_{x}-\omega_{z}$. Instead of use the linear transition basis, namely $\bigl|x\bigr\rangle, \bigl|y\bigr\rangle$ and $\bigl|z\bigr\rangle$, we rather adopt the basis of two circular polarized transition $\bigl|R_{yz}\bigr\rangle$ and $\bigl|L_{yz}\bigr\rangle$ and a linear transition $\bigl|x\bigr\rangle$.

 In the 2D model, each atom interacts only with two perpendicular waveguides, and thus, its effective Hamiltonian can be obtained by summing analogous 1D effective Hamiltonian for a waveguide parallel to x ($\mathcal{H}_{\textrm{eff}}^{x}$) with another one parallel to y ($\mathcal{H}_{\textrm{eff}}^{y}$). Therefore, the effective k-space Hamiltonian in the 2D model is given by $\mathcal{H}_{\textrm{eff}}^{2D}=\sum_{\boldsymbol{k}} \Psi_{\boldsymbol{k}}^{\dagger} (\mathcal{H}_{\textrm{eff}}^{x}+\mathcal{H}_{\textrm{eff}}^{y}) \Psi_{\boldsymbol{k}}$, with $\Psi_{\boldsymbol{k}} = (b_{\boldsymbol{k},R_{yz}} b_{\boldsymbol{k},L_{yz}} b_{\boldsymbol{k},x})^{T}$. Its expression in the vicinity of $k_x,k_y =0$ for $qd=\pi$, obtained in detail in the supplemental material (S.4), reads
\begin{eqnarray}
\mathcal{H}_{\textrm{eff}}^{2D} & = & \frac{\Gamma_{0}}{8}  \sum_{\boldsymbol{k}}  \Psi_{\boldsymbol{k}}^{\dagger}
\left(\begin{array}{ccc}
k_{x}d+\delta_y & 2\delta_x-\delta_y & -\frac{1}{\sqrt{2}}k_{y}d\\
2\delta_x-\delta_y & -k_{x}d+\delta_y & \frac{1}{\sqrt{2}}k_{y}d\\
-\frac{1}{\sqrt{2}}k_{y}d & \frac{1}{\sqrt{2}}k_{y}d & 2\delta_y
\end{array}\right) \Psi_{\boldsymbol{k}} \nonumber \\ 
&  &
\label{eq:H2d-k}
\end{eqnarray}
with corresponding dispersion shown in Fig.~\ref{fig:Fig4}. For the case of degenerate frequencies, $\delta_{x},\delta_{y}=0$, the low-energy spectrum reveals a profile identical to those characteristic to Lieb lattices, with a pair of Dirac cones intersected by a flat band (Fig.~\ref{fig:Fig4} (a)). With a finite detuning in the waveguides parallel to $y-$axis, $\delta_{y}=0.02\Gamma_{0}$, the spectrum shows anisotropy along $k_{y}$ as well as emergence of a gap, as it can be seen in Fig.~\ref{fig:Fig4} (b). Naturally, if a detuning is introduced in the waveguides parallel to $x-$axis the anisotropy will appear along $k_{x}$ instead of $k_{y}$. In Fig.~\ref{fig:Fig4} (c) the both detunings are finite, so that the anisotropy is present both along $x$ and $y$ direction, respectively.

\begin{figure}[t]
    \centerline{\includegraphics[width=3.4in,keepaspectratio]{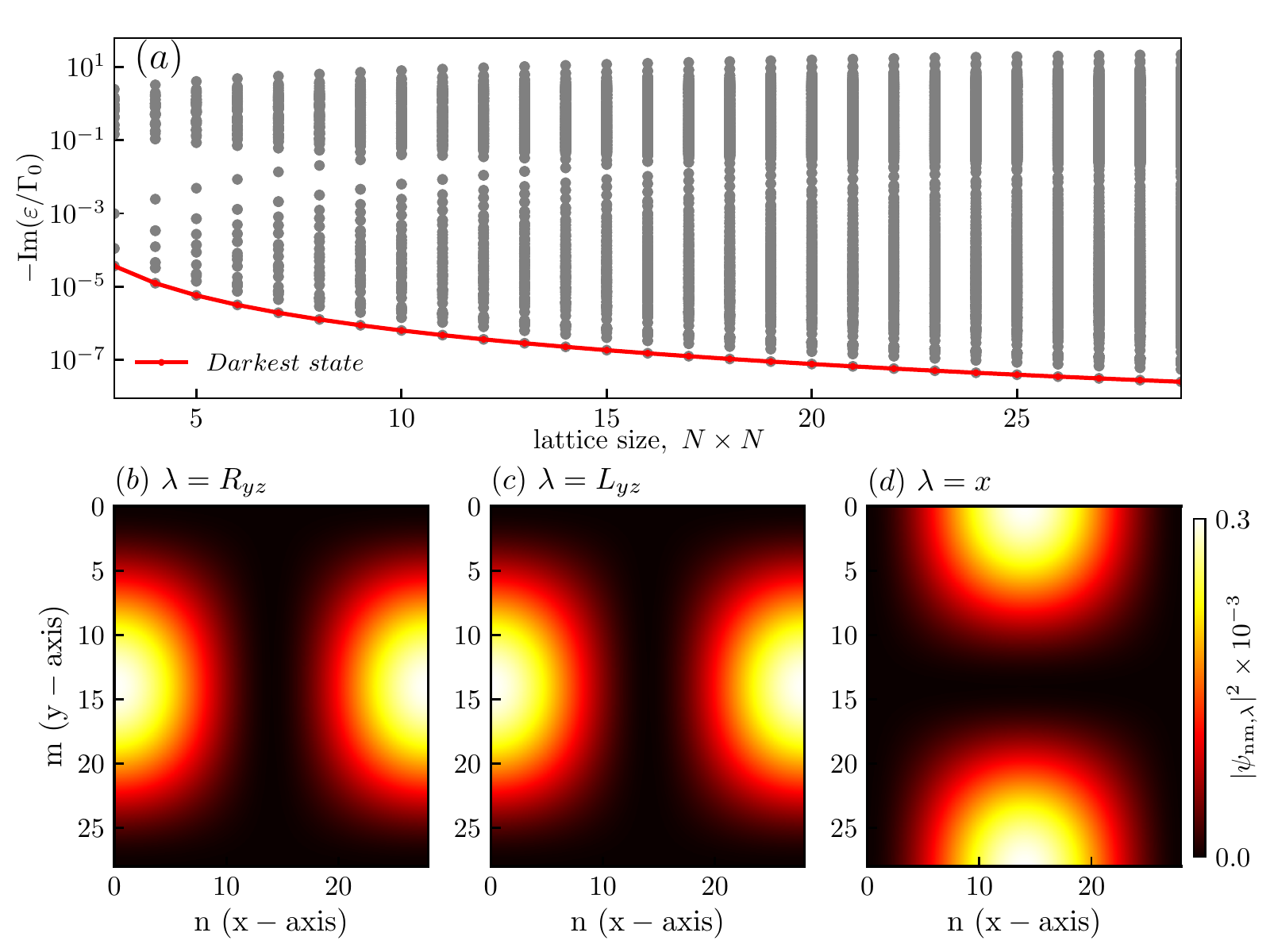}}
    \caption{(a) Radiative decay rate of the complete set of polariton modes as a function of the size of a squared lattice $N \times N$ for $qd=\pi$ with the darkest state highlighted by the red curve. Photonic distribution in the darkest state for $qd=\pi$ with (b) right polarization  ($|\psi_{nm,R_{yz}}|^{2}$), (c) left polarization ($|\psi_{nm,L_{yz}}|^{2}$), and (d) polarization along $x$ ($|\psi_{nm,x}|^{2}$).
    \label{fig:Fig5}}
\end{figure}

Analogously to the k-space description, the effective real-space Hamiltonian is straightforwardly written as
$\mathcal{H}_{\textrm{eff}}^{2D}= \mathcal{H}_{\textrm{eff}}^{x}+\mathcal{H}_{\textrm{eff}}^{y}$,
where
\begin{align}
\mathcal{H}_{\textrm{eff}}^{x} = \sum_{j,n=1}^{N}\delta_{x}\bigl(b_{nj,R_{yz}}^{\dagger}b_{nj,L_{yz}}+\textrm{h.c.}\bigr) -i\frac{\Gamma_{0}}{2}\sum_{m,n=1}^{N} \sum_{j=1}^{N} \nonumber \\
 \times  \Theta(m-n) e^{iqd|m-n|} \bigl(b_{mj,R_{yz}}^{\dagger}b_{nj,R_{yz}}+b_{nj,L_{yz}}^{\dagger}b_{mj,L_{yz}}\bigr),
\label{eq:Heff-xx}
\end{align}
accounts for the set of $N$ interacting qubit chains parallel to $x$, while the chains along $y$ are described by
\begin{align}
\mathcal{H}_{\textrm{eff}}^{y} = \sum_{j,n=1}^{N}\delta_{y}\bigl(b_{nj,R_{zx}}^{\dagger}b_{nj,L_{zx}}+\textrm{h.c.}\bigr) -i\frac{\Gamma_{0}}{2}\sum_{m,n=1}^{N} \sum_{j=1}^{N} \nonumber \\
 \times\Theta(m-n) e^{iqd|m-n|} \bigl(b_{mj,R_{zx}}^{\dagger}b_{nj,R_{zx}}+b_{nj,L_{zx}}^{\dagger}b_{mj,L_{zx}}\bigr).
\label{eq:Heff-yy}
\end{align}
Its diagonalization for $qd=\pi$ with the single occupation basis $\bigl|\Psi\bigr\rangle=\sum_{n,m,\lambda}\psi_{n,m,\lambda}b_{n,m,\lambda}^{\dagger}\bigl|0\bigr\rangle$ leads to the  radiative decay rate and photonic distribution shown in Fig.~\ref{fig:Fig5}.

The imaginary part of the darkest state displayed in Fig.~\ref{fig:Fig5} (a) reveals collective radiative decay rate scaling $\Gamma_{2D} \sim N^{-3}$. Although the 1D chain present a purely chiral propagation for $\delta=0$, the 2D lattice within the condition $\delta_{x},\delta_{y}=0$ does not reveal purely chiral propagation due to the admixture of excitations stemmed from perpendicular waveguides. This admixture can be perceived by the off-diagonal terms proportional to $k_{y}$, such as $b_{\boldsymbol{k},R_{yz}}^{\dagger}b_{\boldsymbol{k},x}$, in Hamiltonian (\ref{eq:H2d-k}). As a consequence, the radiative decay rate scaling $\Gamma_{2D} \sim N^{-3}$ is observed instead of a scaling $\sim N^{-1}$ exclusively observed for purely chiral propagation.
Additionally, the highest group velocity polaritons lie on the Dirac cones and correspond to the superradiant states, while the subradiant states are found quasi-bounded in the flat polariton band.

It can be clearly seen in photonic distribution shown Fig.~\ref{fig:Fig5} (b-d), that right and left circularly polarized eigenmodes in $yx-$plane appear to be more localized at the center of $y-$aligned chains placed at the most left and most right edges of the lattice, while linear polarized modes along $x$ presents similar profile but are found localized at the center of $x-$aligned chains.

\section{Conclusions}
We analyzed dispersions and decay rates of the polariton modes emerging in 1D and 2D arrays of multilevel atoms coupled to chiral waveguides. In particular, it was demonstrated that in 2D case low-energy effective Hamiltonian is similar to those characteristic to the Lieb lattice model and thus generates polariton flat bands, which due to the large density of states may host rich family of strongly interacting multi photonic states.

\begin{acknowledgments}
 \section*{Acknowledgments}
 The work was supported by Russian Science
Foundation (project 20-12-00224) and ITMO 5-100 Program. IAS acknowledges support from Icelandic Research Fund (project "Hybrid polaritonics").
\end{acknowledgments}

\providecommand{\noopsort}[1]{}\providecommand{\singleletter}[1]{#1}%
%

\pagebreak
\widetext
\begin{center}
\textbf{\large Supplemental Material:\\ \bigskip Two-dimensional chiral waveguide quantum electrodynamics: long range qubit
correlations and flat-band dark polaritons}
\end{center}
\setcounter{equation}{0}
\setcounter{figure}{0}
\setcounter{table}{0}
\setcounter{page}{1}
\makeatletter
\renewcommand{\theequation}{S\arabic{equation}}
\renewcommand{\thefigure}{S\arabic{figure}}
\renewcommand{\bibnumfmt}[1]{[S#1]}
\renewcommand{\citenumfont}[1]{S#1}

\section{S.1. Schrieffer-Wolff method}

In this section, we provide the detailed Schrieffer-Wolff approach
performed in the Hamiltonian (\ref{eq:H}) to decouple the atomic subsystem
from the photonic waveguide modes. This procedure consists in perform
a unitary transformation in the Hamiltonian as
\begin{equation}
\mathcal{H}_{\textrm{eff}}=e^{-S}\mathcal{H}e^{S}=H_{0}+H_{v}+\left[H_{0}+H_{v},S\right]+\frac{1}{2}\left[\left[H_{0}+H_{v},S\right],S\right]+...
\end{equation}
in which $H_{0}$ and $H_{v}$ represent the diagonal and off-diagonal
parts of (\ref{eq:H}), and $S$ stands for the anti-Hermitian block off-diagonal
generator. Accordingly to Ref.~\cite{SW-DL2}, the generator chosen under the condition
${\displaystyle [H_{0},S]=-H_{v}}$ removes the off-diagonal term
to the first order and also provides the following block-diagonal
effective Hamiltonian
\begin{equation}
\mathcal{H}_{\textrm{eff}}=H_{0}+\frac{1}{2}\left[H_{v},S\right]+O(g^{3}).
\end{equation}
Hence, by following the procedure proposed in Ref.~\cite{haq2019},
$S$ is obtained as
\begin{eqnarray}
S & = & \frac{g}{\sqrt{2l}}\sum_{n,k}\left[\frac{\sigma_{n,-}^{x}a_{k,L}^{\dagger}}{\omega_{k,L}-\omega_{x}}+\frac{\sigma_{n,-}^{x}a_{k,R}^{\dagger}}{\omega_{k,R}-\omega_{x}}+i\frac{\sigma_{n,-}^{y}a_{k,R}^{\dagger}}{\omega_{k,R}-\omega_{y}}-i\frac{\sigma_{n,-}^{y}a_{k,L}^{\dagger}}{\omega_{k,L}-\omega_{y}}\right]e^{-ikz_{n}}-\textrm{h.c.}
\end{eqnarray}
where the following commutation relations for the raising and lowering
operators have been used
\begin{eqnarray}
\left[\sigma_{n,+}^{\nu},\sigma_{m,+}^{\nu'}\right]=0 &  & \left[\sigma_{n,+}^{\nu}\sigma_{n,-}^{\nu},\sigma_{m,\pm}^{\nu'}\right]=\pm\delta_{nm}\delta_{\nu\nu'}\sigma_{n,\pm}^{\nu}\\
\left[\sigma_{n,-}^{\nu},\sigma_{m,-}^{\nu'}\right]=0 &  & \left[\sigma_{n,-}^{\nu}\sigma_{n,+}^{\nu},\sigma_{m,\pm}^{\nu'}\right]=\mp\delta_{nm}\delta_{\nu\nu'}\sigma_{n,\pm}^{\nu}
\end{eqnarray}
The second order term $\frac{1}{2}\left[H_{v},S\right]$ provides
two decoupled block diagonal Hamiltonians, one describing the waveguide
photons dressed by the atomic chain ($\frac{1}{2}\left[H_{v},S\right]_{\textrm{wg}}$),
and other one characterizing the chain of interacting qubits ($\frac{1}{2}\left[H_{v},S\right]_{\textrm{ch}}$)
that we are interest in. With some straightforward but cumbersome algebra,
the effective Hamiltonian for the qubit chain reads
\begin{eqnarray}
\mathcal{H}_{\textrm{eff}} & = & \sum_{n,\nu}\omega_{\nu}\sigma_{n,+}^{\nu}\sigma_{n,-}^{\nu}+\frac{1}{2}\left[H_{v},S\right]_{\textrm{ch}}\nonumber \\
 & = & \sum_{n,\nu}\omega_{\nu}\sigma_{n,+}^{\nu}\sigma_{n,-}^{\nu}-\frac{g^{2}}{2l}\sum_{k,\nu}\sum_{nm}\left(\frac{1}{\omega_{k,R}-\omega_{\nu}}+\frac{1}{\omega_{k,L}-\omega_{\nu}}\right)e^{ik\left(z_{n}-z_{m}\right)}\sigma_{n,+}^{\nu}\sigma_{m,-}^{\nu}\nonumber \\
 & - & i\frac{g^{2}}{4l}\sum_{k}\sum_{n,m}\left(\frac{1}{\omega_{k,R}-\omega_{x}}-\frac{1}{\omega_{k,L}-\omega_{x}}+\frac{1}{\omega_{k,R}-\omega_{y}}-\frac{1}{\omega_{k,L}-\omega_{y}}\right)e^{ik\left(z_{n}-z_{m}\right)}\left(\sigma_{n,+}^{x}\sigma_{m,-}^{y}-\sigma_{n,+}^{y}\sigma_{m,-}^{x}\right). \label{eq:h0ef}
\end{eqnarray}
Here, we recognize the following hopping amplitudes
\begin{eqnarray}
J_{\nu,R}=-\frac{g^{2}}{4\pi}\begin{cases}
\int_{-\infty}^{\infty}dk\frac{e^{ik|z_{n}-z_{m}|}}{vk-\left(\omega_{\nu}+i0^{+}\right)} & n\geq m\\
\int_{-\infty}^{\infty}dk\frac{e^{-ik|z_{n}-z_{m}|}}{vk-\left(\omega_{\nu}+i0^{+}\right)} & n<m
\end{cases} & \;\textrm{and}\; & J_{\nu,L}=-\frac{g^{2}}{4\pi}\begin{cases}
\int_{-\infty}^{\infty}dk\frac{e^{ik|z_{n}-z_{n'}|}}{-vk-\left(\omega_{\nu}+i0^{+}\right)} & n>m\\
\int_{-\infty}^{\infty}dk\frac{e^{-ik|z_{n}-z_{n'}|}}{-vk-\left(\omega_{\nu}+i0^{+}\right)} & n\leq m
\end{cases}
\end{eqnarray}
wherein the discrete sum is replaced by the one-dimensional continuous
integral $\sum_{k}\rightarrow\int_{-\infty}^{\infty}dk\left(\frac{l}{2\pi}\right)$
and $\omega_{k,R(L)}=\pm vk$. By employing the Cauchy integral formula,
its solutions yield
\begin{eqnarray}
J_{\nu,R}=-\frac{g^{2}}{4\pi}\begin{cases}
\frac{2\pi i}{v}e^{i\frac{\omega_{\nu}}{v}|z_{n}-z_{m}|} & j\geq j'\\
0 & j<j'
\end{cases} & \;\textrm{and}\; & J_{\nu,L}=-\frac{g^{2}}{4\pi}\begin{cases}
0 & j>j'\\
\frac{2\pi i}{v}e^{i\frac{\omega_{\nu}}{v}|z_{n}-z_{m}|} & j\leq j'
\end{cases}
\end{eqnarray}
Hence, equation (\ref{eq:h0ef}) is written as
\begin{eqnarray}
\mathcal{H}_{\textrm{eff}} & = & \sum_{n,\nu}\omega_{\nu}\sigma_{n,+}^{\nu}\sigma_{n,-}^{\nu}-i\frac{\Gamma_{0}}{2}\sum_{\nu}\sum_{n,m}e^{i\frac{\omega_{\nu}}{v}|z_{n}-z_{m}|}\sigma_{n,+}^{\nu}\sigma_{m,-}^{\nu}\nonumber \\
 & - & i\frac{\Gamma_{0}}{4}\sum_{n,m}^{n\neq m}\textrm{sgn}(n-m)\left(e^{i\frac{\omega_{x}}{v}|z_{n}-z_{m}|}+e^{i\frac{\omega_{y}}{v}|z_{n}-z_{m}|}\right)\left(\sigma_{n,+}^{x}\sigma_{m,-}^{y}-\sigma_{n,+}^{y}\sigma_{m,-}^{x}\right),
\end{eqnarray}
in which $\Gamma_{0}=\frac{g^{2}}{v}$. As we are interest only in
the in single-excitation states, the bosonisation procedure for qubits
is employed by replacing $\sigma_{n,+}^{\nu}$ for regular bosonic
operator in the basis of circular polarization with
\begin{eqnarray}
b_{n,x}^{\dagger}=\frac{1}{\sqrt{2}}\left(b_{n,R}^{\dagger}+b_{n,L}^{\dagger}\right) & \;\textrm{and}\; & b_{n,y}^{\dagger}=\frac{i}{\sqrt{2}}\left(b_{n,R}^{\dagger}-b_{n,L}^{\dagger}\right)
\end{eqnarray}
As a result, we get
\begin{eqnarray}
\mathcal{H}_{\textrm{eff}} & = & \sum_{n}\omega\left(b_{n,R}^{\dagger}b_{n,R}+b_{n,L}^{\dagger}b_{n,L}\right)-i\frac{\Gamma_{0}}{4}\sum_{n,m}\left(e^{i\frac{\omega_{x}d}{v}|n-m|}+e^{i\frac{\omega_{y}d}{v}|n-m|}\right)\left(b_{n,R}^{\dagger}b_{m,R}+b_{n,L}^{\dagger}b_{m,L}\right)\nonumber\\
 & + & \sum_{n}\delta\left(b_{n,R}^{\dagger}b_{n,L}+b_{n,L}^{\dagger}b_{n,R}\right)-i\frac{\Gamma_{0}}{4}\sum_{n,m}\left(e^{i\frac{\omega_{x}d}{v}|n-m|}-e^{i\frac{\omega_{y}d}{v}|n-m|}\right)\left(b_{n,R}^{\dagger}b_{m,L}+b_{n,L}^{\dagger}b_{m,R}\right)\nonumber\\
 & - & \frac{i\Gamma_{0}}{4}\sum_{n,m}^{n\neq m}\textrm{sgn}(n-m)\left(e^{i\frac{\omega_{x}d}{v}|n-m|}+e^{i\frac{\omega_{y}d}{v}|n-m|}\right)\left(b_{n,R}^{\dagger}b_{m,R}-b_{n,L}^{\dagger}b_{m,L}\right)
\end{eqnarray}
Within the Markovian approximation ($\delta,g\ll\omega$), where Schrieffer-Wolff
approach is valid, the effective Hamiltonian can be reduced to
\begin{eqnarray}
\mathcal{H}_{\textrm{eff}} & = & \sum_{n}\omega\left(b_{n,R}^{\dagger}b_{n,R}+b_{n,L}^{\dagger}b_{n,L}\right)+\sum_{n}\delta\left(b_{n,R}^{\dagger}b_{n,L}+b_{n,L}^{\dagger}b_{n,R}\right) \nonumber\\
 & - & i\frac{\Gamma_{0}}{2}\sum_{m,n}\Theta(m-n)e^{iqd|m-n|}\left(b_{m,R}^{\dagger}b_{n,R}+b_{n,L}^{\dagger}b_{m,L}\right)
\end{eqnarray}
in which $q=\omega/v$ stands for the wavevector of the photon, and
$\Theta(m-n)$ is the Heaviside step function defined within the half-maximum
convention, i.e., $\Theta(0)=1/2$.

\section{S.2. One-dimensional k-space effective Hamiltonian derivation}

In this section, we derive the k-space effective Hamiltonian of the one-dimensional system by considering the infinite lattice limit and performing a Fourier transform in the Hamiltonian (\ref{eq:Heff}) of the main text. Upon defining the Fourier transform of the bosonic operators as
\begin{eqnarray}
b_{m,\lambda}^{\dagger}=\frac{1}{\sqrt{l}}\sum_{k}e^{-ikdm}b_{k,\lambda}^{\dagger} & \ \textrm{for} \ & \lambda = R,L,
\end{eqnarray}
the Hamiltonian (\ref{eq:Heff}) for an infinite array of atoms is rewritten as
\begin{eqnarray}
\mathcal{H}_{\textrm{eff}}(k) & = & -i\frac{\Gamma_{0}}{2l}\sum_{m,n=-\infty}^{\infty}\sum_{kk'}\Theta(m-n)e^{iqd|m-n|}\bigl(e^{-id(km-k'n)}b_{k,R}^{\dagger}b_{k',R}+e^{id(km-k'n)}b_{k,L}^{\dagger}b_{k',L}\bigr)\nonumber \\
 & + & \frac{1}{l} \sum_{n=-\infty}^{\infty}\sum_{kk'}\delta\bigl(e^{-id(k-k')n}b_{k,R}^{\dagger}b_{k',L}+\textrm{h.c.}\bigr).
\end{eqnarray}
Next, we substitute the sum oven $n$ and $m$ to $j=m-n$ and $j'=m+n$ in the first right-hand side term and recognize Dirac delta distribution $l\delta_{k,k'} =\sum_{j'}e^{id(k'-k)\frac{j'}{2}}$ to find
\begin{eqnarray}
\mathcal{H}_{\textrm{eff}}(k) & = & -i\frac{\Gamma_{0}}{2}\sum_{kk'}\sum_{j=-\infty}^{\infty}\delta_{k,k'}\Theta(j)e^{iqd|j|}\bigl(e^{-id(k+k')\frac{j}{2}}b_{k,R}^{\dagger}b_{k',R}+e^{id(k+k')\frac{j}{2}}b_{k,L}^{\dagger}b_{k',L}\bigr)\nonumber \\
 & + & \sum_{kk'}\delta\bigl(\delta_{k,k'}b_{k,R}^{\dagger}b_{k',L}+\textrm{h.c.}\bigr).
\end{eqnarray}
After some algebra, we obtain
\begin{eqnarray}
\mathcal{H}_{\textrm{eff}}(k) & = & -i\frac{\Gamma_{0}}{2}\sum_{k}\Bigl(\frac{1}{2}+\sum_{j=1}^{\infty}e^{id(q-k)j}\Bigr)b_{k,R}^{\dagger}b_{k,R}-i\frac{\Gamma_{0}}{2}\sum_{k}\Bigl(\frac{1}{2}+\sum_{j=1}^{\infty}e^{id(q+k)j}\Bigr)b_{k,L}^{\dagger}b_{k,L}\nonumber\\
 & + & \sum_{k}\delta\bigl(b_{k,R}^{\dagger}b_{k,L}+\textrm{h.c.}\bigr),
\end{eqnarray}
where the solution of the sum in $j$ is given by
\begin{eqnarray}
\sum_{j=1}^{\infty}e^{id(q\pm k)j} & = & -\frac{1}{2}+\frac{i}{2}\cot\Bigl(\frac{qd\pm kd}{2}\Bigr).
\end{eqnarray}
Therefore, the effective Hamiltonian in k-space in matrix form reads
\begin{eqnarray}
\mathcal{H}_{\textrm{eff}}(k) & = & \sum_{k}\left(\begin{array}{cc}
b_{k,R}^{\dagger} & b_{k,L}^{\dagger}\end{array}\right)\left(\begin{array}{cc}
\frac{\Gamma_{0}}{4}\cot\Bigl(\frac{qd-kd}{2}\Bigr) & \delta\\
\delta & \frac{\Gamma_{0}}{4}\cot\Bigl(\frac{qd+kd}{2}\Bigr)
\end{array}\right)\left(\begin{array}{c}
b_{k,R}\\
b_{k,L}
\end{array}\right),
\end{eqnarray}
with corresponding eigenvalues $\varepsilon_{k}$ for different phases $qd$ shown in Fig.~\ref{fig:Fig2}.

\section{S.3. Transfer matrix}

For the considered one-dimensional setup the transfer matrix over
the single period of the structure composed by a layer of thickness
$d$ with a scatterer in the middle (at the point $z=0$) reads
\begin{eqnarray}
\mathcal{T}=\frac{1}{t}\left(\begin{array}{cc}
\left(t^{2}-r^{2}\right)e^{iqd} & r\\
-r & e^{-iqd}
\end{array}\right) & ; & \ \ t=1+r,\label{eq:T0}
\end{eqnarray}
in which $r$ and $t$ respectively stand for amplitude reflection
and transmission coefficients of the scatterer. The relation between
the transfer matrix with the incident and reflected electric fields
amplitudes $E'_{\pm}$ and $E_{\pm}$ is given by
\begin{eqnarray}
\left(\begin{array}{c}
E'_{+}\\
E'_{-}
\end{array}\right) & = & \mathcal{T}\left(\begin{array}{c}
E_{+}\\
E_{-}
\end{array}\right).
\end{eqnarray}
As the right and left propagating waves are circularly polarized,
the electric fields amplitudes are written as $E_{R,L}=\hat{R}^{-1}E_{\pm}$
with
\begin{eqnarray}
\hat{R}=\frac{1}{\sqrt{2}}\left(\begin{array}{cc}
1 & i\\
1 & -i
\end{array}\right) \ \ & \textrm{and} & \ \ \hat{R}^{-1}=\frac{1}{\sqrt{2}}\left(\begin{array}{cc}
1 & 1\\
-i & i
\end{array}\right).
\end{eqnarray}
Hence, the pair of levels at frequencies $\omega_{x}$ and $\omega_{y}$
of the scatterer, written in the circularly polarized basis, is
characterized by
\begin{eqnarray}
\hat{h}=\hat{R}\hat{h}_{0}\hat{R}^{-1} & = & \left(\begin{array}{cc}
\frac{i\Gamma_{0}}{\omega_{x}-\varepsilon_{k}-i\Gamma} & 0\\
0 & \frac{i\Gamma_{0}}{\omega_{y}-\varepsilon_{k}-i\Gamma}
\end{array}\right)=\hat{R}\left(\begin{array}{cc}
\frac{i\Gamma_{0}}{\omega_{x}-\varepsilon_{k}} & 0\\
0 & \frac{i\Gamma_{0}}{\omega_{y}-\varepsilon_{k}}
\end{array}\right)\hat{R}^{-1}=i\Gamma_{0}\left(\begin{array}{cc}
\frac{2(\omega-\varepsilon_{k})}{(\omega-\varepsilon_{k})^{2}-\delta^{2}} & \frac{-2\delta}{(\omega-\varepsilon_{k})^{2}-\delta^{2}}\\
\frac{-2\delta}{(\omega-\varepsilon_{k})^{2}-\delta^{2}} & \frac{2(\omega-\varepsilon_{k})}{(\omega-\varepsilon_{k})^{2}-\delta^{2}}
\end{array}\right),
\end{eqnarray}
wherein we considered bound states ($\Gamma=0$). The corresponding
Green's functions in reciprocal and real spaces for the counter propagating
modes with dispersion $\omega_{k}=\pm vk$ are respectively given
as
\begin{eqnarray}
\hat{G}_{0}(k)=\left(\begin{array}{cc}
\frac{1}{kv-\omega_{k}} & 0\\
0 & \frac{-1}{kv+\omega_{k}}
\end{array}\right); &  & \ \ \hat{G}_{0}(z)=\left(\begin{array}{cc}
\Theta(z)e^{ik|z|} & 0\\
0 & \Theta(-z)e^{-ik|z|}
\end{array}\right).
\end{eqnarray}
Within the Green's function formalism~\cite{economou-2013}, the transfer matrix can be
written as 
\begin{eqnarray}
\mathcal{T}(z) & = & \hat{h}+\hat{h}\hat{G}_{0}(z)\hat{h}+\hat{h}\hat{G}_{0}(z)\hat{h}\hat{G}_{0}(z)\hat{h}+...\nonumber \\
 & = & \frac{\hat{h}}{(\hat{I}-\hat{G}_{0}(z)\hat{h})},\label{eq:T_z}
\end{eqnarray}
with the amplitude reflection and transmission coefficients given
by
\begin{eqnarray}
t=1+\left\{ \hat{h}\times\left[\hat{I}-\hat{G}_{0}(0)\hat{h}\right]^{-1}\right\} _{11} \ \ & \textrm{and} & \ \
r = \left\{ \hat{h}\times\left[\hat{I}-\hat{G}_{0}(0)\hat{h}\right]^{-1}\right\} _{12}.
\end{eqnarray}
By employing this approach to our setup, the coefficients are obtained
as
\begin{eqnarray}
t=\frac{\left(\omega-\varepsilon_{k}\right)^{2}-\delta^{2}+\Gamma_{0}^{2}/4}{\left(\left(\omega-\varepsilon_{k}\right)-i\Gamma_{0}^{2}/2\right)^{2}-\delta^{2}} \ \  & \textrm{and} \ \  & r=-\frac{i\delta\Gamma_{0}}{\left(\left(\omega-\varepsilon_{k}\right)-i\Gamma_{0}^{2}/2\right)^{2}-\delta^{2}}.
\end{eqnarray}
Therefore, the transfer matrix (\ref{eq:T0}) becomes
\begin{eqnarray}
\mathcal{T} & = & \frac{1}{\left(\omega-\varepsilon_{k}\right)^{2}-\delta^{2}+\Gamma_{0}^{2}/4}\left(\begin{array}{cc}
e^{iqd}\left[\left(\left(\omega-\varepsilon_{k}\right)+i\Gamma_{0}^{2}/2\right)^{2}-\delta^{2}\right] & i\delta\Gamma_{0}\\
-i\delta\Gamma_{0} & e^{-iqd}\left[\left(\left(\omega-\varepsilon_{k}\right)-i\Gamma_{0}^{2}/2\right)^{2}-\delta^{2}\right]
\end{array}\right)
\end{eqnarray}
and the dispersion equation for the eigenmodes, defined as $\cos(kd)=\frac{1}{2}\mathrm{Tr}(\mathcal{T})$, is obtained as
\begin{eqnarray}
\cos(kd) & = & \cos(qd)\left[\frac{\left(\omega-\varepsilon_{k}\right)^{2}-\delta^{2}-\Gamma_{0}^{2}/4}{\left(\omega-\varepsilon_{k}\right)^{2}-\delta^{2}+\Gamma_{0}^{2}/4}\right]-\sin(qd)\left[\frac{\left(\omega-\varepsilon_{k}\right)\Gamma_{0}}{\left(\omega-\varepsilon_{k}\right)^{2}-\delta^{2}+\Gamma_{0}^{2}/4}\right].
\end{eqnarray}

\section{S.4. Derivation of two-dimensional k-space effective Hamiltonian}

As discussed in the main text, the atoms assume a four-level configuration with dipole excited states $\bigl|R_{yz}\bigr\rangle$, $\bigl|L_{yz}\bigr\rangle$, and $\bigl|x\bigr\rangle$, in which the detuning frequency between the corresponding energy levels are $2\delta_{x}= \omega_{y}-\omega_{z}$ and $2\delta_{y}= \omega_{x}-\omega_{z}$. In the 2D waveguide network each atom interacts only with two waveguides, and thus, its effective Hamiltonian in k-space is given by the sum of analogous 1D effective Hamiltonian, one align in x given by
\begin{eqnarray}
\mathcal{H}_{\textrm{eff}}^{x} & = & \frac{\Gamma_{0}}{4}\sum_{\boldsymbol{k}}\left[\cot\left(\frac{qd-k_{x}d}{2}\right)b_{\boldsymbol{k},Ryz}^{\dagger}b_{\boldsymbol{k},Ryz}+\cot\left(\frac{qd+k_{x}d}{2}\right)b_{\boldsymbol{k},Lyz}^{\dagger}b_{\boldsymbol{k},Lyz}+\delta_{x}\left(b_{\boldsymbol{k},Ryz}^{\dagger}b_{\boldsymbol{k},Lyz}+\textrm{h.c.}\right)\right] \label{eq:Heff-x}
\end{eqnarray}
and another one aligned in y direction with
\begin{eqnarray}
\mathcal{H}_{\textrm{eff}}^{y} & = & \frac{\Gamma_{0}}{4}\sum_{\boldsymbol{k}}\left[\cot\left(\frac{qd-k_{y}d}{2}\right)b_{\boldsymbol{k},Rzx}^{\dagger}b_{\boldsymbol{k},Rzx}+\cot\left(\frac{qd+k_{y}d}{2}\right)b_{\boldsymbol{k},Lzx}^{\dagger}b_{\boldsymbol{k},Lzx}+\delta_{y}\left(b_{\boldsymbol{k},Rzx}^{\dagger}b_{\boldsymbol{k},Lzx}+\textrm{h.c.}\right)\right] \label{eq:Heff-y}
\end{eqnarray}
in which the bosonic operators $b_{\boldsymbol{k},R_{yz}}$ and $b_{\boldsymbol{k},L_{yz}}$ in the basis of $\Psi_{\boldsymbol{k}} = (b_{\boldsymbol{k},R_{yz}} b_{\boldsymbol{k},L_{yz}} b_{\boldsymbol{k},x})^{T}$ are writen as
\begin{eqnarray}
\begin{cases}
b_{\boldsymbol{k},Rxz} & =\frac{1}{\sqrt{2}}\left(b_{\boldsymbol{k},z}-ib_{\boldsymbol{k},x}\right)\\
b_{\boldsymbol{k},Lxz} & =\frac{1}{\sqrt{2}}\left(b_{\boldsymbol{k},z}+ib_{\boldsymbol{k},x}\right)
\end{cases} & \rightarrow & \begin{cases}
b_{\boldsymbol{k},Rxz} & =\frac{i}{2}\left(b_{\boldsymbol{k},Ryz}-b_{\boldsymbol{k},Lyz}-\sqrt{2}b_{\boldsymbol{k},x}\right)\\
b_{\boldsymbol{k},Lxz} & =\frac{i}{2}\left(b_{\boldsymbol{k},Ryz}-b_{\boldsymbol{k},Lyz}+\sqrt{2}b_{\boldsymbol{k},x}\right)
\end{cases} \label{eq:bo}
\end{eqnarray}
Performing the basis change in the Hamiltonian (\ref{eq:Heff-y}) using the definition (\ref{eq:bo}), we find
\begin{eqnarray}
\mathcal{H}_{\textrm{eff}}^{y} & = & -\frac{\Gamma_{0}}{4}\sum_{\boldsymbol{k}}\left[\frac{\frac{1}{2}\sin\left(qd\right)}{\cos\left(qd\right)-\cos\left(k_{y}d\right)}\right]\left(b_{\boldsymbol{k},Ryz}^{\dagger}b_{\boldsymbol{k},Ryz}-b_{\boldsymbol{k},Ryz}^{\dagger}b_{\boldsymbol{k},Lyz}-b_{\boldsymbol{k},Lyz}^{\dagger}b_{\boldsymbol{k},Ryz}+b_{\boldsymbol{k},Lyz}^{\dagger}b_{\boldsymbol{k},Lyz}+2b_{\boldsymbol{k},x}^{\dagger}b_{\boldsymbol{k},x}\right)\nonumber\\
 & - & \frac{\Gamma_{0}}{4}\sum_{\boldsymbol{k}}\left[\frac{\frac{1}{2}\sin\left(k_{y}d\right)}{\cos\left(qd\right)-\cos\left(k_{y}d\right)}\right]\left(-\sqrt{2}b_{\boldsymbol{k},Ryz}^{\dagger}b_{\boldsymbol{k},x}+\sqrt{2}b_{\boldsymbol{k},Lyz}^{\dagger}b_{\boldsymbol{k},x}-\sqrt{2}b_{\boldsymbol{k},x}^{\dagger}b_{\boldsymbol{k},Ryz}+\sqrt{2}b_{\boldsymbol{k},x}^{\dagger}b_{\boldsymbol{k},Lyz}\right)\nonumber\\
 & + & \frac{\delta_{y}}{2}\sum_{\boldsymbol{k}}\left(b_{\boldsymbol{k},Ryz}^{\dagger}b_{\boldsymbol{k},Ryz}-b_{\boldsymbol{k},Ryz}^{\dagger}b_{\boldsymbol{k},Lyz}-b_{\boldsymbol{k},Lyz}^{\dagger}b_{\boldsymbol{k},Ryz}+b_{\boldsymbol{k},Lyz}^{\dagger}b_{\boldsymbol{k},Lyz}-2b_{\boldsymbol{k},x}^{\dagger}b_{\boldsymbol{k},x}\right)
\end{eqnarray}
Therefore, the 2D effective Hamiltonian  $\mathcal{H}_{\textrm{eff}}^{2D}=\sum_{\boldsymbol{k}} \Psi_{\boldsymbol{k}}^{\dagger} (\mathcal{H}_{\textrm{eff}}^{x}+\mathcal{H}_{\textrm{eff}}^{y}) \Psi_{\boldsymbol{k}}$ reads
\begin{eqnarray}
\mathcal{H}_{\textrm{eff}}^{2D} & = & \frac{\Gamma_{0}}{4}  \sum_{\boldsymbol{k}}  \Psi_{\boldsymbol{k}}^{\dagger}
 \left(\begin{array}{ccc}
\cot\left(\frac{qd-k_{x}d}{2}\right)+\frac{\frac{1}{2}\sin(qd)}{\cos(qd)-\cos(k_{y}d)} & \frac{-\frac{1}{2}\sin(qd)}{\cos(qd)-\cos(k_{y}d)} & \frac{-\frac{1}{\sqrt{2}}\sin(k_{y}d)}{\cos(qd)-\cos(k_{y}d)}\\
\frac{-\frac{1}{2}\sin(qd)}{\cos(qd)-\cos(k_{y}d)} & \cot\left(\frac{qd+k_{x}d}{2}\right)+\frac{\frac{1}{2}\sin(qd)}{\cos(qd)-\cos(k_{y}d)} & \frac{\frac{1}{\sqrt{2}}\sin(k_{y}d)}{\cos(qd)-\cos(k_{y}d)}\\
\frac{-\frac{1}{\sqrt{2}}\sin(k_{y}d)}{\cos(qd)-\cos(k_{y}d)} & \frac{\frac{1}{\sqrt{2}}\sin(k_{y}d)}{\cos(qd)-\cos(k_{y}d)} & \frac{\sin(qd)}{\cos(qd)-\cos(k_{y}d)}
\end{array}\right)  \Psi_{\boldsymbol{k}}
 \nonumber \\
 & + & 
 \sum_{\boldsymbol{k}} \Psi_{\boldsymbol{k}}^{\dagger}
\left(\begin{array}{ccc}
\frac{\delta_y}{2}  & \delta_x-\frac{\delta_y}{2} & 0\\
\delta_x-\frac{\delta_y}{2} & \frac{\delta_y}{2} & 0\\
0 & 0 & \delta_y
\end{array}\right) \Psi_{\boldsymbol{k}}. 
\label{eq:H0}
\end{eqnarray}
The dispersion relations of the 1D model obtained via Schrieffer-Wolff are in good agreement with the exact dispersion around $k_{x},k_{y}=0$ only for phases $qd=\pi/2$ and $qd=\pi$, as can be seen in Fig.~\ref{fig:Fig2} of the main text. However, the solutions for $qd=\pi/2$  are uninteresting and then disregarded. The series expansions of $\mathcal{H}_{\textrm{eff}}^{2D}(\boldsymbol{k})$ in the vicinity of $k_{x},k_{y}=0$ for $qd=\pi$ is leads to
\begin{eqnarray}
\mathcal{H}_{\textrm{eff}}^{2D} & = & \frac{\Gamma_{0}}{8}  \sum_{\boldsymbol{k}}  \Psi_{\boldsymbol{k}}^{\dagger}
\left(\begin{array}{ccc}
k_{x}d+\delta_y & 2\delta_x-\delta_y & -\frac{1}{\sqrt{2}}k_{y}d\\
2\delta_x-\delta_y & -k_{x}d+\delta_y & \frac{1}{\sqrt{2}}k_{y}d\\
-\frac{1}{\sqrt{2}}k_{y}d & \frac{1}{\sqrt{2}}k_{y}d & 2\delta_y
\end{array}\right) \Psi_{\boldsymbol{k}}
\label{eq:H0}
\end{eqnarray}
with its respective eigenmodes depending on the detunings $\delta_x$ and $\delta_y$ displayed in Fig.~\ref{fig:Fig4} of the main text.

\providecommand{\noopsort}[1]{}\providecommand{\singleletter}[1]{#1}%

\end{document}